\documentclass[3p,times]{elsarticle}

\usepackage{amssymb}

\usepackage[figuresright]{rotating}

\usepackage{graphicx}
\usepackage{dcolumn}
\usepackage{bm}

\newcommand {\apgt} {\ {\raise-.5ex\hbox{$\buildrel>\over\sim$}}\ }
\newcommand {\aplt} {\ {\raise-.5ex\hbox{$\buildrel<\over\sim$}}\ }

\begin{document}
\begin{frontmatter}




\title{{First Principles Phase Diagram Calculation for
the 2D TMD system $WS_{2}-WTe_{2}$.}}


\author{B. P. Burton }
\address{Materials Measurement Laboratory,
National Institute of Standards and Technology (NIST),
Gaithersburg, MD 20899, USA; benjamin.burton@nist.gov}

\begin{abstract}
First principles phase diagram calculations, that included
van der Waals interactions, were performed for the bulk
transition metal dichalcogenide
system $(1-X) \cdot WS_{2} - (X) \cdot WTe_{2}$.
To obtain a converged phase diagram, a series of cluster expansion
calculations were performed with increasing numbers of structure-energies,
($N_{Str}$) up to $N_{Str}=435$, used to fit the cluster expansion Hamiltonian.
All calculated formation energies are positive and all ground-state
analyses predict that formation energies for supercells with 16 or
fewer anion sites are positive; but when  $\approx 150 N_{Str} \leq 376$,
false ordered ground-states are predicted. With $N_{Str} \geq 399$, only a
miscibility gap is predicted, but one with dramatic asymmetry opposite to
what one expects from size-effect considerations; i.e. the calculations
predict more solubility on the small-ion S-rich side of the diagram and
less on the large-ion Te-rich side.  This occurs because S-rich low-energy
metastable ordered configurations have lower energies than their Te-rich counterparts.

\end{abstract}


\end{frontmatter}



\section{Introduction}

There is great interest in two-dimensional (2D)
transition metal dichalcogenide (TMD) materials
$MX_{2}$, where M = Mo, W, Nb, Re, etc. and X = S, Se, or Te.  \cite{Wang,Ganatra2014}.  
Currently, interest is focused on applications such as:
band-gap engineering \cite{Kang,Kutana}; 
nano-electronic devices \cite{Ganatra2014,Radisavljevic2011,Das2013,Wang2012}; 
photovoltaic devices \cite{Jariwala2014,Fontana2013}; 
valleytronics applications \cite{Zeng2012,Mak2012};
2D building blocks for electronic heterostructures \cite{2D}; and
as sensors \cite{Sensor}.  

The bulk 2H crystal structure ($P6_{3}/mmc$~ space group) has AB-stacking of 
three-atom-thick 2D-layers that are bonded by van der Waals forces. 
Hence van der Waals forces influence bulk and multilayer 
phase relations and therefore anion order-disorder and/or phase separation 
in TMD solid solutions.  The results presented below, for bulk $WS_{2}-WTe_{2}$,  
imply that van der Waals interactions may strongly affect phase stabilities, 
either between adjacent layers in bulk or few-layer samples, or between 
monolayers and substrates.   

Previous work on bulk $(1-X) \cdot MoS_{2}-(X) \cdot MoTe_{2}$ \cite{Burton2016} 
predicted two entropy stabilized incommensurate phases at $X\approx0.46$, 
and this work was done to see if a similar prediction applies to the structurally analogous
$(1-X) \cdot WS _{2} -(X) \cdot WTe _{2}$~ system. In the $WS_{2}-WTe_{2}$~ system, 
however, only a miscibility gap is predicted, but a very large number of formation energy 
calculations, $N_{Str} \aplt 400$, is required to suppress false ground-states (GS). 
Also, the asymmetry of the calculated phase diagram is the opposite of what one expects 
from a size-effect argument; typically there is more solubility of the smaller ion in 
larger-ion-rich solutions (more S-solubility in Te-rich solutions) than vice versa;
$R_{S}$=1.84 \AA; $R_{Te}$=2.21 \AA). \cite{Radii}

\section{Methodology}

\subsection{Total Energy Calculations}

Total structure energies, $\Delta E_{Str}$~ were calculated 
for fully relaxed $WS_{2}$, $WTe_{2}$~ and for 433 
W$_{m+n}$(S$_{m}$Te$_{n}$)$_{2}$~ supercells. 
The Vienna $ab~initio$~ simulation program 
(VASP, version 5.3.3 ~\cite{Kresse1993,Disclaimer}) 
was used for all density-functional theory (DFT) calculations, with
projector augmented waves (PAW) and a generalized 
gradient approximation (GGA) for exchange energies. 
Electronic degrees of freedom were optimized with a conjugate gradient
algorithm. Valence electron configurations were: 
W\_pv$~5p^{5}6s5d$; S\_$s^{2}p^{4}$; Te\_$s^{2}p^{4}$.   
Van der Waals interactions modeled with the
non-local correlation functional of Klimes $et$~ $al.$ \cite{Klimes} 
A 500 eV cutoff-energy was used in the "high precision" option, 
which converges {\it absolute}~ energies to 
within a few meV/mol (a few tenths of a kJ/mol of exchangeable 
S- and Te-anions). Precision is at least an order of
magnitude better. Residual forces of order 0.02 eV or less were typical.

\subsection{The Cluster Expansion Hamiltonian}

Cluster expansion Hamiltonians (CEH) \cite{Sanchez1984},
were fit to sets of 71, 253, 295, 399, and 435 formation energies, 
$\Delta E_{f}$, solid dots (green online) in Figs. \ref{fg:PD71}a-\ref{fg:PD435}a:

\begin{equation}
\Delta E_{f} = (E_{Str} - mE_{WS_{2}} - nE_{WTe_{2}})/(2(m+n))
\end{equation}

\noindent
Here: $E_{Str}$~ is the total energy of the W$_{m+n}$(S$_{m}$Te$_{n}$)$_{2}$~ supercell;
$E_{WS{2}}$  is the energy/mol of WS$_{2}$;
$E_{WTe{2}}$ is the energy/mol of WTe$_{2}$.

Fittings of the CEHs were performed with the Alloy Theoretic Automated Toolkit (ATAT)
\cite{Disclaimer,Axel2002a,Axel2002b,Axel2002c} which automates most CEH construction
tasks \cite{Axel2002b}. 

\section{Results}

\subsection{Ground-State Analyses}

Filled circles (green online) in Figs. \ref{fg:PD71}a - \ref{fg:PD435}a 
indicate values of $\Delta E_{f}$~ that were calculated with the VASP package,
i.e. $\Delta E_{VASP}$.
Large open squares in Figs. \ref{fg:PD71}a - - \ref{fg:PD435}a (red online) 
indicate the CEH-fit to $\Delta E_{VASP}$.  Smaller open squares 
($\Delta E_{GS}$; blue online) indicate the results of a ground-state (GS) analyses
that included  all ordered configurations with 16 or fewer anion sites, 
151,023 structures.  Calculated values for cross validation scores, (CV)$^2$, 
and the numbers of structures, $N_{Str}$, are plotted on the figures.
   
Additional GS analyses were performed by Monte-Carlo (MC)
simulations at fixed bulk compositions, via decreasing temperature (T) scans 
down to T=0. The 0K $\Delta E_{f}$~ values from these calculations 
are plotted as solid- (predicted stable) or open-diamonds (metastable; blue online) 
in Figs. \ref{fg:PD253}a and \ref{fg:PD295}a; and as small filled down-facing triangles 
(blue online) in Figs. \ref{fg:PD399}a and \ref{fg:PD435}a. Because the calculated 
formation energies for the ordered configurations in 
Figs. \ref{fg:PD253}a and \ref{fg:PD295}a are negative, 
they constitute (falsely) predicted large-cell ordered-GS. 
If their formation energies are positive they can be regarded
as low-energy microstructures. Note that these formation energies from MC-simulations
are always upper bounds, because MC-simulations don't yield $perfectly$~ ordered
simulation boxes.

\subsection{Phase Diagram Calculations}

\begin{figure}[!htbp]
\begin{center}
\vspace{-0.25in}
\includegraphics[width=12.cm,angle=0]{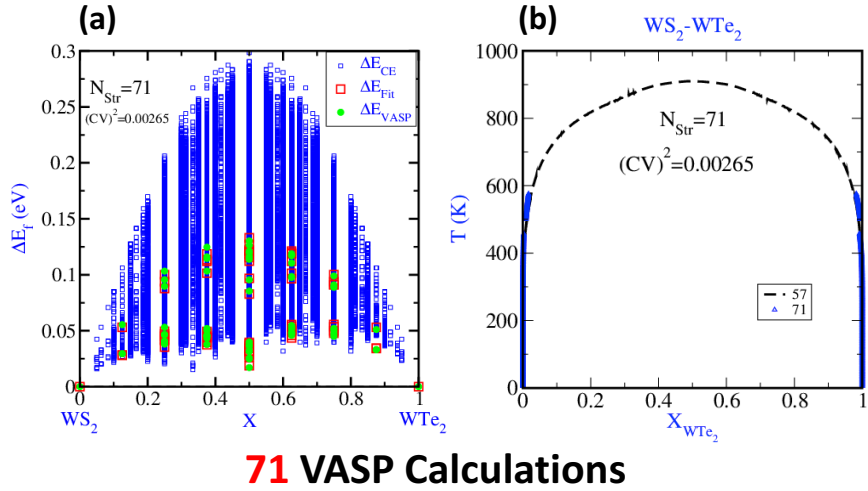}
\end{center}
\vspace{-0.30in}
\caption{For $N_{Str}=71$: (a) Ground-State analysis; (b) calculated phase diagram.
In (a): $\Delta E_{VASP}$~ filled circles (green online); 
$\Delta E_{Fit}$~  large open squares, (red online) is the CE-fit to the DFT set; 
$\Delta E_{CE}$~ smaller open squares, (blue online) are the CE-based ground-state analysis;
All $\Delta E_{f}$\textgreater$0$~ implies that there are no ordered
GS, with 16 or fewer anion sites, and suggests that the phase diagram will
have a miscibility gap. Note the small cross-validation score, (CV)$^2=0.00265$, 
which suggests a very good CEH-fit, and in (b) the near absence of 
asymmetry in the miscibility gap.
} 
\label{fg:PD71}
\end{figure}

\begin{figure}[!htbp]
\begin{center}
\vspace{-0.25in}
\includegraphics[width=12.cm,angle=0]{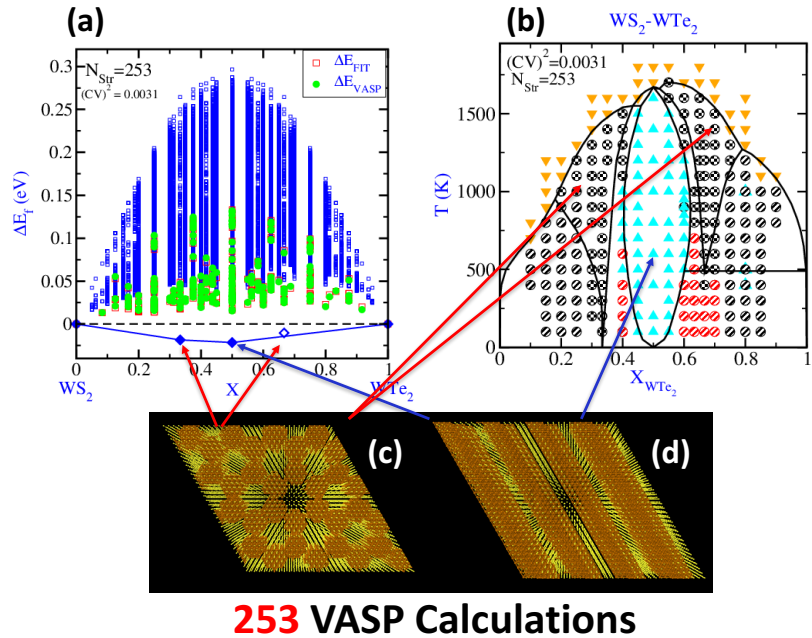}
\end{center}
\vspace{-0.30in}
\caption{For $N_{Str}=253$: (a) ground-state analysis; (b) calculated phase diagram.
Filled diamond symbols in (a) indicate 
predicted GS structures as shown in the MC-snapshots of: 
(c) honeycomb structure at $X=1/3$, and (d) a striped-phase at $X=1/2$. 
The open diamond symbol at $X=2/3$~ indicates a low-energy 
metastable honeycomb-ordered structure.  
Additional symbols in (b): 
large filled down-pointing triangles (orange online) indicate disorder;
up-pointing triangles (cyan online) indicate a layer structure (d);
large checkered circles (red online) indicate a honeycomb structure (c);
and striped circles (black and red online) indicate two-phase,
assemblages, ordered plus disordered or two ordered phases.
} 
\label{fg:PD253}
\end{figure}

\begin{figure}[!htbp]
\begin{center}
\vspace{-0.25in}
\includegraphics[width=12.cm,angle=0]{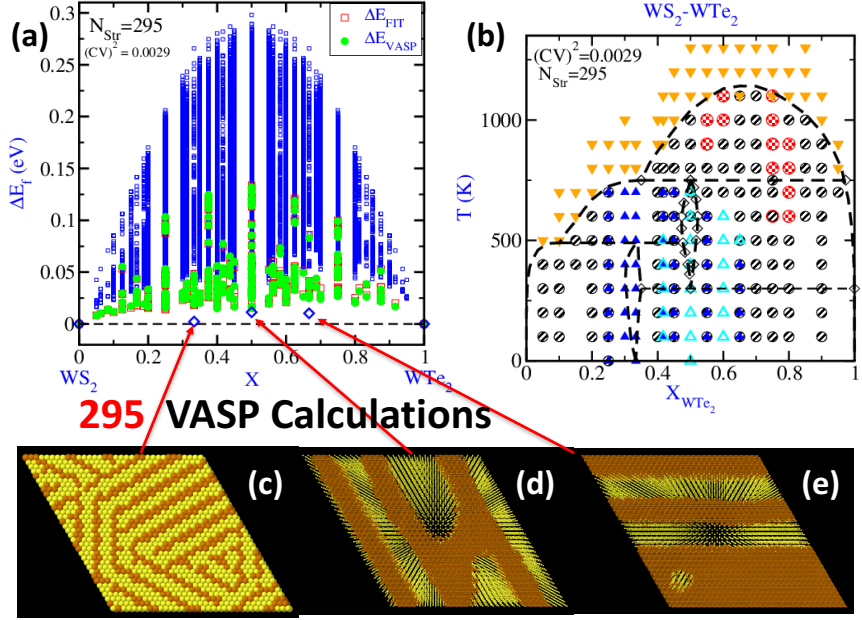}
\end{center}
\vspace{-0.30in}
\caption{For $N_{Str}=295$: (a) ground-state analysis; (b) calculated phase diagram.
Open diamonds in (a) (blue online) indicate: (c) an 
ordered structure at $X=1/3$; and low-energy, mostly striped, microstructures 
at $X = 1/2$~ and $X = 2/3$. Note however, that the $X=1/3$- and $X=1/2$-phases
appear to be stable at elevated temperatures.
} 
\label{fg:PD295}
\end{figure}

\begin{figure}[!htbp]
\begin{center}
\vspace{-0.25in}
\includegraphics[width=12.cm,angle=0]{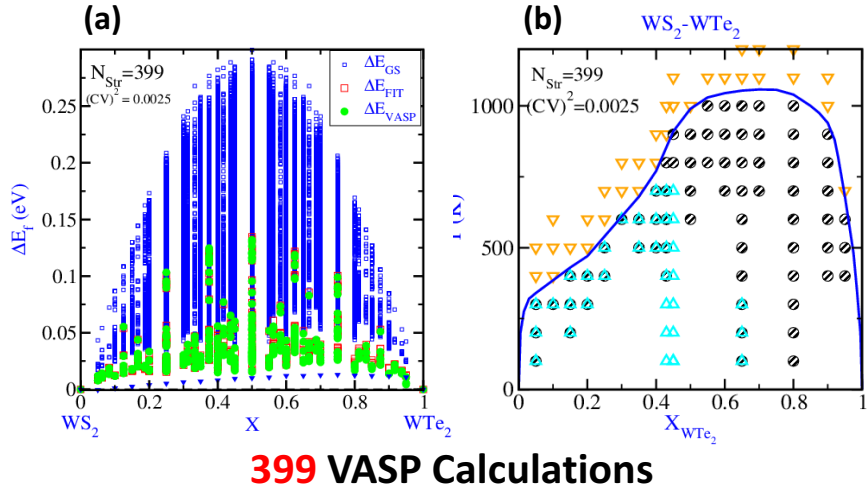}
\end{center}
\vspace{-0.30in}
\caption{For $N_{Str}=399$: (a) ground-state analysis; (b) calculated phase diagram.
Small down-pointing triangles in (a) are $\Delta E_{CE}$~
values for MC-simulation T-scans from a low-T value to T=0. 
Note the asymmetry in these values.
Additional symbols in (b) have the same meanings as in Fig. \ref{fg:PD253}b.
} 
\label{fg:PD399}
\end{figure}

\begin{figure}[!htbp]
\begin{center}
\vspace{-0.25in}
\includegraphics[width=12.cm,angle=0]{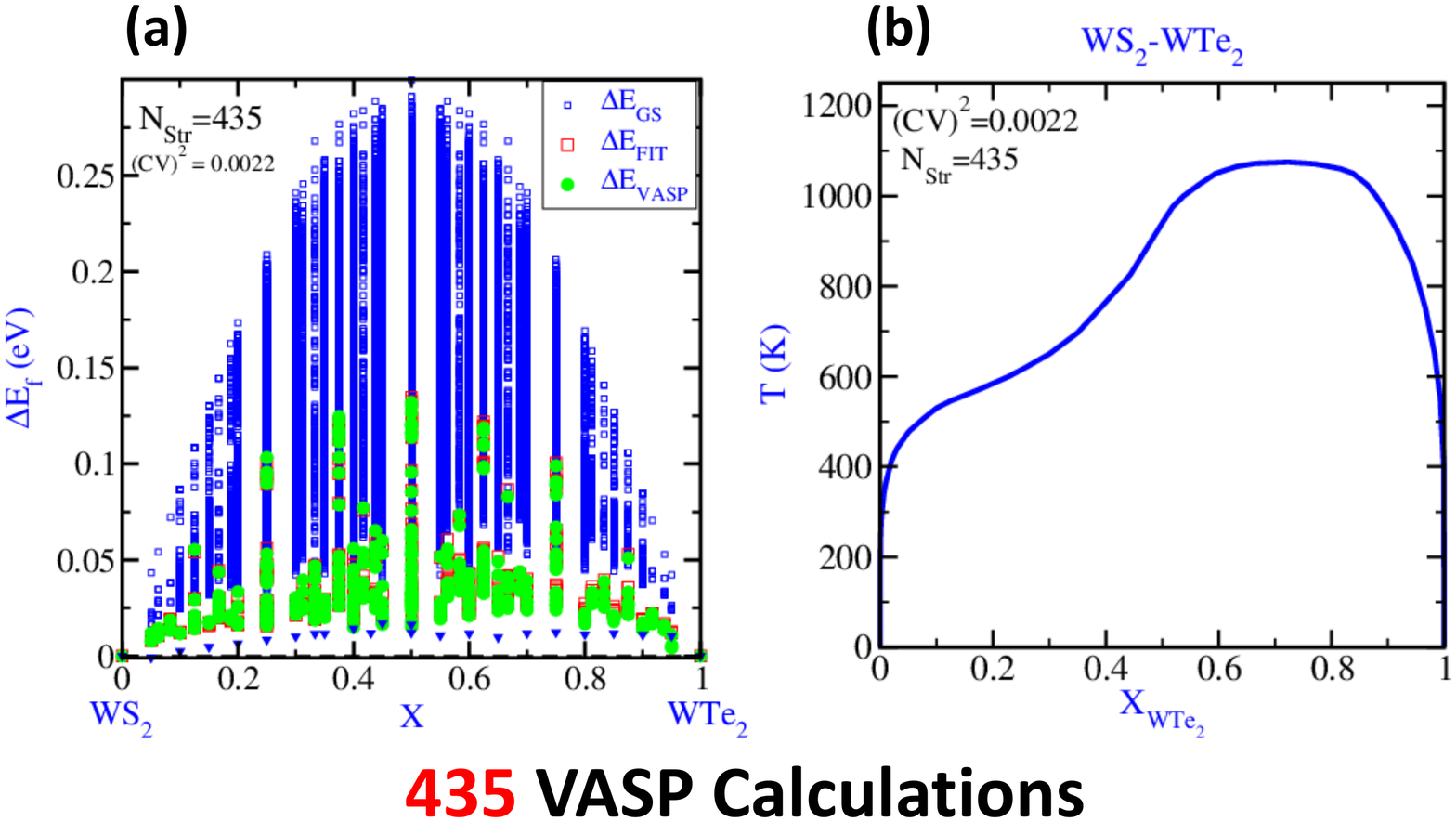}
\end{center}
\vspace{-0.30in}
\caption{For $N_{Str}=435$: (a) ground-state analysis; (b) calculated phase diagram.
Down-pointing triangles in (a) are $\Delta E_{CE}$~
values for MC-simulation T-scans from a low-T value to T=0. Note the
asymmetry in these values, which is opposite to what one expects from
a size-effect argument. Compare the nearly symmetric miscibility gap in 
Fig. \ref{fg:PD71}b with the dramatic asymmetries of 
Figs. \ref {fg:PD399}b and \ref {fg:PD435}b.
} 
\label{fg:PD435}
\end{figure}

First principles phase diagram calculations that were performed
with the ATAT package \cite{Axel2002a,Axel2002b,Axel2002c} are plotted
in Figs. \ref{fg:PD71}b-\ref{fg:PD435}b. Additional symbols on 
Figs. \ref{fg:PD253}b, \ref{fg:PD295}b, and \ref{fg:PD399}b are 
used to indicate various phase fields that were identified, 
by visual inspections of MC-snapshots: 
large filled down-pointing triangles (orange online) indicate disorder;
up-pointing triangles (cyan online) indicate a layer structure 
(e.g. Fig. \ref{fg:PD253}d); 
large checkered circles (red online) indicate a honeycomb structure (Fig. \ref{fg:PD253}c);
and striped circles (black and red online) indicate two-phase, 
assemblages, ordered plus disordered or two ordered phases.

\section{Discussion}

One expects that fitting CEHs to larger and larger sets of $\Delta E_{VASP}$~
ultimately leads to a converged result for 
the calculated phase diagram. The results presented
here indicate that the fits with $N_{Str}$ = 71, 253, 295, and 376 (not shown) 
are not sufficient because: false GS are predicted, typically at $X=1/3$~ and
$X=1/2$; and qualitatively different phase diagrams are predicted with each
increase in $N_{Str}$.  Standard ground-state analyses for the sets with 
$N_{Str}$ = 253, 295, and 376 (not shown) predicted no ordered GS with 16 
or fewer anion sites, but MC T-scans down to T=0K, predicted false GS based 
on unit cells with more than 16 anion sites. The diagrams for $N_{Str}$ = 399 or 435 are 
essentially identical, and may represent a converged result.  One can, however, 
never rule out the possibility that a fit based on $N_{Str} > 435$, might 
yield a different result.  

Two generalizations apply to all calculated phase diagrams for models
with $ 150 \aplt N_{Str} \leq 376$:
(1) When false GS are predicted, they are always in the S-rich bulk composition
range $0 \aplt X \leq 0.5$; (2) The range range $0.5 \apgt X \leq 1.0$~
is dominated by phase separation at T$ \aplt 1050K$. (1) above indicates
that low-energy ordered configurations on the S-rich side of the system
drive the asymmetry of phase separation that is noted in (2).

Kang et al. \cite{Kang} performed first principles phase diagram calculations
(with ATAT; $N_{Str} \approx 40$) for monolayer $WS_{2} - WTe_{2}$, 
and reported a phase diagram with its' consolute point at 
$(X,T) \approx (0.55,680K)$; i.e. without the dramatic asymmetry 
exhibited in Figs. \ref {fg:PD399} and \ref {fg:PD435} where 
$(X,T) \approx (0.7,1075K)$.

\section{Conclusions}

A CEH-fit to at least $N_{Str} \approx 400$~ is required to calculate
a realistic phase diagram for the $WS_{2} - WTe_{2}$~ TMD
system.  Low cross-validation scores, and routine GS analyses are not 
sufficient for systems such as TMDs because very low-energy metastable 
ordered states imply that an apparently well-fit CEH can predict false GS
phases. It is likely that the $WS_{2}-WTe_{2}$~ system has a highly asymmetric 
miscibility gap as shown in Fig. \ref {fg:PD435},
and that the predicted asymmetry is driven by low-energy metastable 
ordered states on the S-rich side of the system.


\section{ACKNOWLEDGEMENTS}
This work was supported by NIST-MGI.


\begin{thebibliography}{999}
\bibitem{Disclaimer} 
Reference to specific software packages does not imply
a NIST endorsement.
\end{thebibliography}
\end{document}